\documentclass[aip,
rsi,
 amsmath,amssymb,
 reprint,%
]{revtex4-2}

\usepackage{graphicx}
\usepackage{dcolumn}
\usepackage{bm}

\usepackage[utf8]{inputenc}
\usepackage[T1]{fontenc}
\usepackage{mathptmx}
\usepackage{etoolbox}
\usepackage{gensymb}
\usepackage{xcolor}

\makeatletter
\def\@email#1#2{%
 \endgroup
 \patchcmd{\titleblock@produce}
  {\frontmatter@RRAPformat}
  {\frontmatter@RRAPformat{\produce@RRAP{*#1\href{mailto:#2}{#2}}}\frontmatter@RRAPformat}
  {}{}
}%
\makeatother
\begin{document}


\title[A new beamline for REBEL and STRIPE]{A new beamline for Resonant Excitation of Beams with Electromagnetic fields and Lasers (REBEL) and Stopping and Trapping of Radioactive Isotopes for Precision Experiments (STRIPE)}
\author{Phillip \surname{Imgram}}
\affiliation{Instituut voor Kern- en Stalingsfysica, KU Leuven, Leuven, Belgium}
\email{phillip.imgram@kuleuven.be}

\author{Dinko \surname{Atanasov}}
\affiliation{Belgian Nuclear Research Centre, SCK CEN, Mol, Belgium}%

\author{Michail \surname{Athanasakis-Kaklamanakis}}
\altaffiliation[Current address: ]{JILA, National Institute of Science and Technology and University of Colorado, Boulder, CO, USA}
\affiliation{Instituut voor Kern- en Stalingsfysica, KU Leuven, Leuven, Belgium}

\author{Paul \surname{Van den Bergh}}
\affiliation{Instituut voor Kern- en Stalingsfysica, KU Leuven, Leuven, Belgium}

\author{Tobias \surname{Christen}}
\affiliation{Instituut voor Kern- en Stalingsfysica, KU Leuven, Leuven, Belgium}

\author{Ruben \surname{de Groote}}%
\affiliation{Instituut voor Kern- en Stalingsfysica, KU Leuven, Leuven, Belgium}

\author{\'Agota \surname{Koszor\'us}}
\affiliation{Instituut voor Kern- en Stalingsfysica, KU Leuven, Leuven, Belgium}
\affiliation{Belgian Nuclear Research Centre, SCK CEN, Mol, Belgium}%

\author{Gerda \surname{Neyens}}%
\affiliation{Instituut voor Kern- en Stalingsfysica, KU Leuven, Leuven, Belgium}

\author{Stefanos \surname{Pelonis}}
\affiliation{Instituut voor Kern- en Stalingsfysica, KU Leuven, Leuven, Belgium}

\date{\today}

\begin{abstract}
We present two newly constructed experimental setups -- REBEL (Resonant Excitation of Beams with Electromagnetic fields and Lasers) and STRIPE (Stopping and Trapping of Radioactive Isotopes for Precision Experiments) -- integrated into a single offline beamline at KU Leuven. REBEL is designed for collinear laser spectroscopy of ion bunches following isobaric separation with a multireflection time-of-flight mass spectrometer, enabling high-sensitivity measurements of mass-selected fast-ion beams. In contrast, STRIPE focuses on the deceleration, trapping, and laser cooling of ions in a segmented linear Paul trap, optimized for long interrogation times and precision spectroscopy. The shared infrastructure features stable high-voltage operation ($<10$~ppm), modular vacuum sections, and a fast-beam switchyard to route ions to either experiment. Initial results include a mass-resolving power of $R \approx 12900$ in REBEL and successful ion trapping and laser cooling of ions with a kinetic energy of 10\,keV in STRIPE, with improved performance achieved using a frequency-modulated cooling laser. This dual-system platform enables the development and benchmarking of advanced spectroscopy and trapping techniques and is compatible with future operation at radioactive ion beam facilities.
\end{abstract}

\maketitle

\section{Introduction}
\begin{figure*}[!t]
\includegraphics[width=\linewidth]{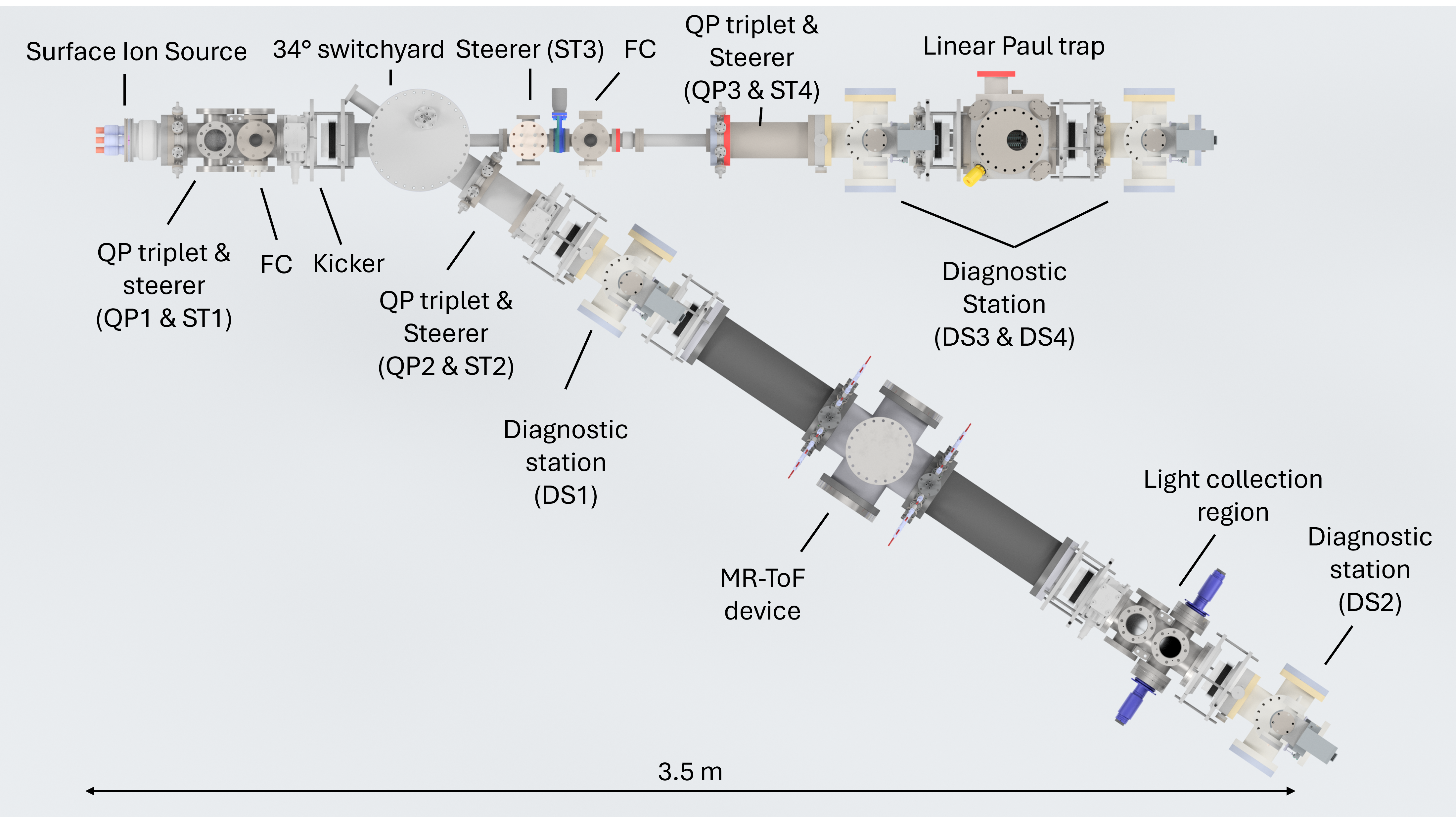}
\caption{\label{fig:rebel_overview}Overview of the Resonant Excitation of Beams with Electromagnetic fields and Lasers (REBEL) beamline. Ions created in an ion-source area, consisting of a surface ion source, a quadrupole (QP) triplet, steerer, Faraday Cup (FC) and a kicker electrode, can either be directed into a collinear laser spectroscopy (CLS) beamline by a 34\degree~switchyard, or passed into a linear Paul trap including a electrostatic deceleration section. The CLS beamline features a second QP triplet and steerer, two stations with beam diagnostics, a multi-reflection time-of-flight (MR-ToF) device and an optical detection region. Each part is described in the text.}
\end{figure*}

The size and deformation of radioactive nuclei are important observables, used to test our understanding of the fundamental interactions interlinked in atomic nuclei. For several decades, the workhorses in the investigation of these properties in ground states and long-lived isomeric states of nuclei have been collinear laser spectroscopy (CLS) and resonant ionization spectroscopy (RIS) and their merger (CRIS). Their importance, achievements, and ever-ongoing technical improvements are documented in a series of review articles \cite{blaum13, Campbell2016, Neugart2017, Yang2023, Koszorus2024}. Most CLS and CRIS setups are naturally located at radioactive ion beam (RIB) facilities. However, offline beamlines have proven their usefulness in the past by providing reference values for online measurements \cite{König20COALAReview, Koenig2020, Ratajcyk2024}, performing precise laser spectroscopy for atomic structure investigations \cite{Borghs1985, childs1992overview, Imgram19, Mueller2020, Imgram2023}, and developing the techniques further \cite{Poulsen1982, Childs1988, Krämer18, Koenig2024_HV}. These developments are timely as new radioactive ion beam laboratories are expected to start operation in the coming years, allowing the implementation of novel techniques, e.g. the future ISOL@MYRRHA facility in Belgium \cite{nuPec_LRP_2025}.
\\
Laser spectroscopy in the collinear geometry is typically limited to fast (Einstein coefficient $>10^6$ Hz) transitions, usually optical dipole-allowed transitions, because of the short interaction time of a few $\mu$s, while plenty of interesting and new physics can be found in slower transitions, as they can be probed with higher accuracy. Examples of new information can be to e.g. investigate nuclear magnetic octupole moments \cite{Schwartz1955, deGroote2024} or hyperfine anomalies from accurate hyperfine structure studies, or to search for physics beyond the standard model \cite{Berengut2020} using isotope-shift spectroscopy. One route to addressing such transitions using long interaction timescales is to decelerate the (radioactive) ions from their initial beam energy (which in most low-energy radioactive ion beam laboratories is 10-60\,keV), down to a few eV, in order to then trap them in a linear Paul trap. Here, the ions can be laser-cooled to temperatures of mK and high-precision laser spectroscopy can be performed. In cases where the spectroscopy should be performed on the neutral atom instead, atom trapping or thermal-beam experiments could be considered. In this paper, however, we restrict our attention to charged species. 
\\
A second way in which current laser spectroscopy experiments could be improved is in their ability to cope with contaminants, i.e. stable or radioactive species in the ion beam other than the ion of interest. Such contaminants typically lead to unwanted background, e.g. as they emit unwanted photons through collisions, or are ionized through non-resonant ionization processes. Developing the means to remove these contaminants from the ion beam, prior to the spectroscopy step, would improve the signal-to-background ratio and would enable spectroscopy on increasingly challenging cases.
\\
In order to develop and commission a trap suited to laser-cooling of radioactive ions, and also to test and improve the CLS technique in combination with a multi-reflection time-of-flight (MR-ToF) device, we constructed a multifunctional offline beamline at the Institute for Nuclear Physics of KU Leuven. The beamline, shown in Fig.\,\ref{fig:rebel_overview}, is described in Sec.\,\ref{sec:beamline} and first commissioning results are presented in Sec.\,\ref{sec:results}. Some ideas for future developments and measurements are outlined in Sec.\,\ref{sec:conclusion}.
\section{The beamline}
\label{sec:beamline}
The new beamline has four dedicated sections, namely an ion source, a switchyard, and two extension arms to serve the REBEL (Resonant Excitation of Beams with Electromagnetic fields and Lasers) and STRIPE (Stopping and Trapping of Radioactive Isotopes for Precision Experiments) experimental setups. The full beamline volume is evacuated to ultra-high vacuum ($p \leq 1 \cdot 10^{-8}$\,mbar) and is split in five sectors by means of gate valves. This enables us to perform quick interventions of single sections, such as a refill or change of the ion source, while preserving the vacuum conditions of the other parts. 
\\
A special feature of REBEL is the combination of an MR-ToF device with a subsequent light collection region, which enables laser spectroscopy of mass-separated ion bunches. This feature (MR-ToF + light collection region) is useful for investigating systematics and possible sensitivity gain for CLS, e.g., by manipulating trapped ions with laser light \cite{Maier2023} in preparation for subsequent laser spectroscopy.
\subsection{Ion source}
\label{sec:ion_source}
\begin{figure}[!t]
\includegraphics[width=0.8\linewidth]{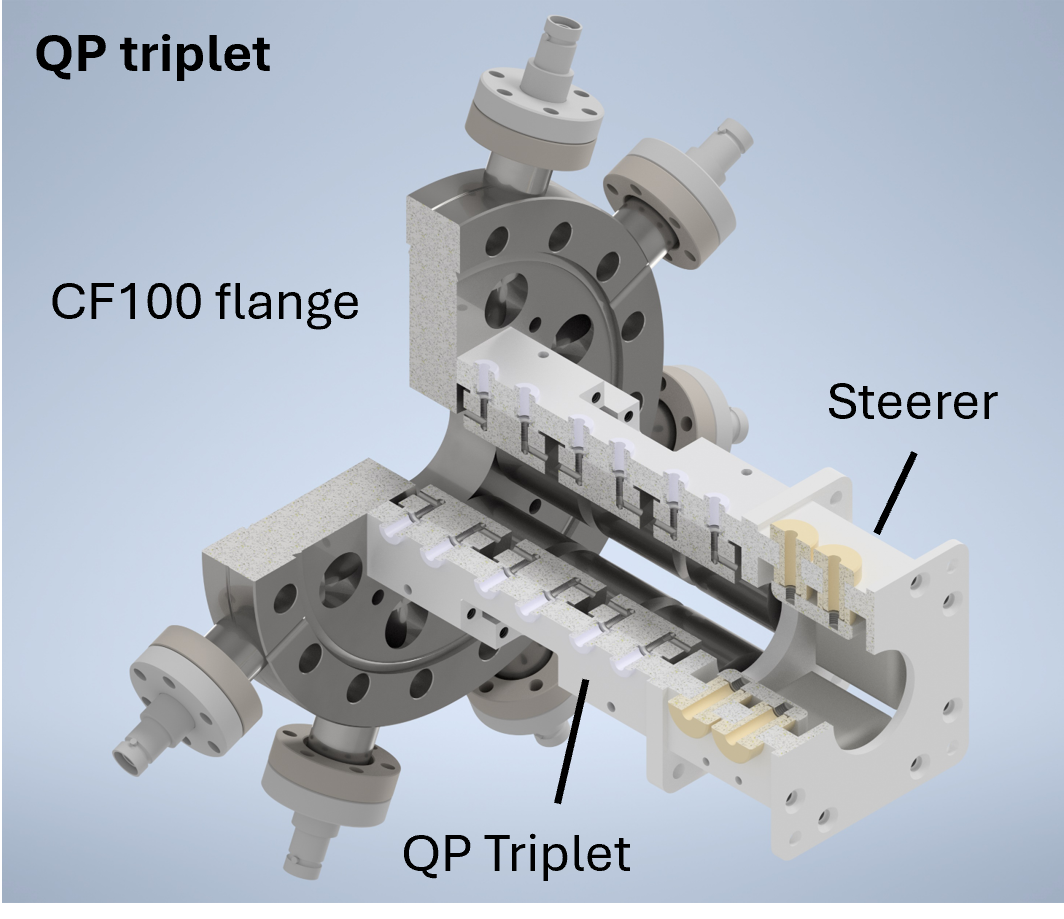}
\caption{\label{fig:QP_triplet}CAD model of the used quadrupole triplets with attached $x$-$y$-steerer in three-quarter view. The ion optics are mounted to a double-sided CF100 flange with 8 SHV feedthroughs. The QP triplet consists of three identical QP section with a length of 30\,mm and a distance of 5 mm between the sections. The electrodes have a radius $r = 20$\,mm and a surface-distance radius $r_0 = 12$\,mm. The diameter of the opening aperture is 30\,mm.}
\end{figure}

The ions are produced on a positive high-voltage (HV) potential of up to 10\,kV and subsequently accelerated towards the ground potential into the beamline, forming a continuous ion beam. Here, a stable absolute potential with respect to ground is necessary for precise CLS measurements and to reliably trap the ions in the Paul trap since this potential defines the start energy of the ions. We achieve a short- (few s) and long-term (few hours) relative and absolute precision with $< 10$\,parts-per-million (ppm) by using an active feedback loop. The HV supply (Heinzinger PNC10000-6 pos) is connected to the ground terminal of a DAC card (Labjack T4) on a high-voltage platform. The DAC card then adds a variable low voltage (0-10\,V) and feeds it to the platform. A software feedback loop adapts the voltage to an HV measurement with a precision high voltage divider from TU Darmstadt \cite{Koenig2024_HV} and a precision voltmeter (Keithley DMM7510). More details about this method can be found in Ref.\,\onlinecite{Mueller2020} and Ref.\,\onlinecite{Koenig2024_HV}.
\\
The HV cage was designed to be large enough to fit a variety of different ion sources (surface ionization, ablation, liquid-metal, \ldots). During the first tests and commissioning of the new beamline, a surface ion source similar to the one described in Ref.\,\onlinecite{König20COALAReview} is used. It can reliably produce ions of alkaline, alkaline earth metals, and even some lanthanides with low ionization potential by resistively heating a graphite crucible to 2100\,\degree C (50\,A, 3\,V). An attached Pierce-shaped electrode helps to form a well-collimated beam during the acceleration towards the ground electrode formed by a double-sided flange holding the subsequent electrostatic quadrupole (QP) triplet which is used to collimate and shape the beam. This results in a beam with a typical normalized emittance $\epsilon_\mathrm{norm} = 490(120)\,\pi\,\mathrm{mm\,mrad}\,\sqrt{\mathrm{eV}}$ as derived in Ref.\,\onlinecite{König20COALAReview}.
\\
The QP triplet shown in Fig.,\ref{fig:QP_triplet} consists of three identical QP sections with a length of 30\,mm and a distance of 5 mm between the sections. The electrodes have a radius $r = 20$\,mm and a surface-distance radius $r_0 = 12$\,mm. The diameter of the opening aperture is 30\,mm. A first $x$-$y$-steerer is attached to the QP triplet, and a subsequent Faraday Cup (FC) can be used to measure the produced ions directly after the ion source.
\\
The surface ion source can produce a continuous Sr$^+$ ion beam of up to 10\,nA. For the presently described applications, an ion current of 300\,pA or less was sufficient, which preserves the sample inside the ion source. The source is filled with 3 mg metal chunks of Sr, requiring refilling after 20 days of continuous use.
For both the MR-ToF and the linear Paul trap, beams with a time structure are necessary. For this, we chop the continuous ion beam into bunches with one kicker electrode and a fast high-voltage switch (Behlke GHTS60A). The shortest bunches achieved with this technique have a temporal length of 1\,$\mu$s. 
\subsection{Switchyard}
\begin{figure}[!t]
\includegraphics[width=0.75\linewidth]{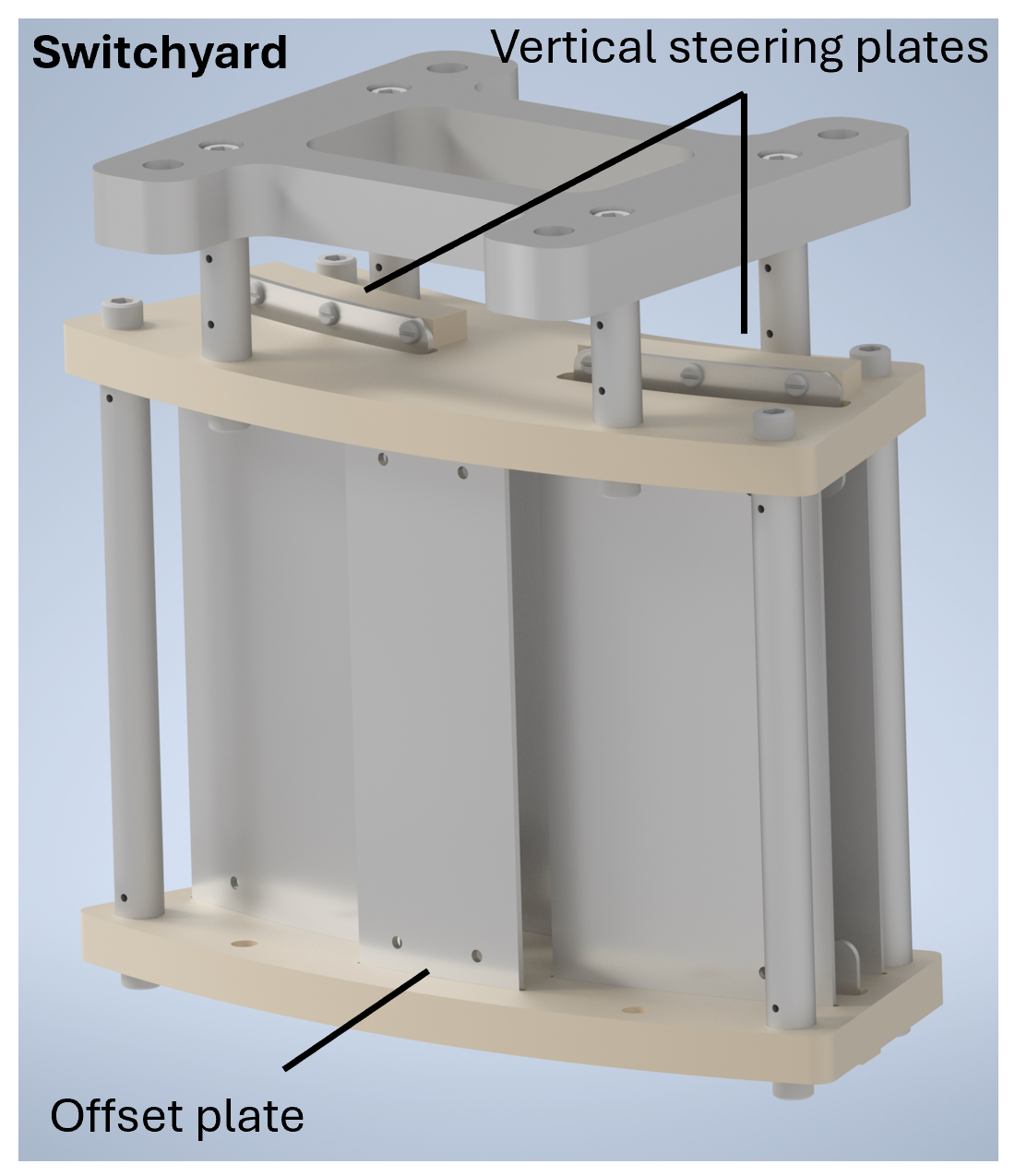}
\caption{\label{fig:diverter}CAD model of the 34\degree switchyard. A segmented outer plate enables the deflection of the ion beam into the CLS section or the transmission towards the linear Paul trap.}
\end{figure}

In order to share resources, REBEL and STRIPE use the same ion source area through a 34\degree~switchyard, which has originally been designed by the University of Manchester for the Collinear Resonant Ionization Spectroscopy (CRIS) experiment at ISOLDE/CERN \cite{Cocolios2013, Vernon2020}. A CAD model of the assembly is shown in Fig.\,\ref{fig:diverter}. The switchyard consists of two bent plates with a distance of 30\,mm, where the outer plate is segmented into three parts. The middle segment has an offset from the other two segments, which creates a gap of 12.5\,mm where either a laser beam or an ion beam can pass in the straight direction. The arc length of the inner plate is 180.6\,mm. Additionally, four small plates in between the large bending plates can be used to steer the ion beam slightly vertically. All plates are mounted in a CNC-machined peek isolator which is directly mounted to the CF250 top flange of the custom vacuum chamber. The chamber has two CF100 ports on the main bending direction and two CF40 ports on the straight axes. While one CF40 port is used for the linear Paul trap, the other axis is used to couple a laser beam into the CLS beamline through an optical view port. In order to bend the ion beam with 3\,keV of kinetic energy towards the CLS beamline, we typically apply a voltage gradient of 500\,V to the main bending plates with HV modules from ISEG-HV.
\\
Steering electrodes with symmetric bending plates induce a beam focus which is different for the horizontal and vertical axis. To compensate for this aberration, a second QP triplet is placed directly after the port towards the CLS beamline. This enables a recollimation of the ion beam. This QP triplet also includes an $x$-$y$-steerer which can be used together with the first steerer to align the ion beam on the main beam axis defined by two iris diaphragms mounted to the beam diagnostic stations.
\subsection{Diagnostic stations}
Identical diagnostic stations (DS), which contain different detectors to characterize the ion beam, are placed at different locations along the beamline. Currently, the DS include homemade Faraday Cups (FC) with a repeller connected to a picoammeter (9103, rbd instruments), and MagneTOF (ETP Ion Detect) devices for single-ion detection. Both detectors are mounted on a ladder and can be inserted through a motorized linear Z-stage (Metallic Flex GmbH). This holder is placed on top of a standard CF100 six-way cross. This allows us to upgrade the stations in the future with a beam observation system on the horizontal axis of the cross.
\\
In front of each DS, an iris diaphragm (Sahm Feinwerktechnik GmbH) similar to the one described in Ref.\,\onlinecite{Klink2024} is mounted on a double-sided CF100 flange. It can be opened and closed through a linear feedthrough. In our case, the irises serve two main purposes: They define the central beam axis, along which all ion optics are aligned, and they help to estimate the size of the ion beam. In the CLS section, they additionally enable the precise superposition of laser and ion beam and help with the reduction of laser-induced stray light. As described in Ref.\,\onlinecite{Klink2024}, they could also be used as a simple FC when they are electrically isolated from the holder flange.

\subsection{MR-ToF device}
Isobar separation with the MR-ToF technique is achieved by trapping a mono-energetic ion beam between two electrostatic mirrors and letting it reflect multiple times, effectively extending its flight path. Each revolution of the ions increases the total flight path, which leads to a spatial and temporal separation of isotopes with different masses. This enables mass-resolving powers $R = m/\Delta m$ beyond $10^5$ in a few tens of ms \cite{Schury2013, Dickel2015, Jesch2015, Wienholtz2015}. 
This principle can be used not only for isobar separation \cite{Yavor2015, Hirsh2016}, but also for mass spectrometry \cite{Plass2013, Rosenbusch2018, Nies2023, Moon2024, Nies2025} and lifetime investigations \cite{Knoll1999, Wolf2016}. Thus, MR-ToF devices have emerged as a versatile tool to complement broad physics communities. Currently, ongoing efforts focus on coupling such a device to the CLS of trapped ions \cite{Maier2023}.
\\
The multi-reflection time-of-flight mass spectrometer discussed here is a direct copy of the design of Schlaich \textit{et al.}\cite{Schlaich2024}. The original purpose of this MR-ToF device is an enhanced isobar separation of laser-ionized atoms for the CRIS experiment at ISOLDE/CERN \cite{Vernon2022}. This will further enhance the experimental sensitivity by removing isobaric contamination. Due to the limited time for new technical implementations at CRIS, the MR-ToF device was assembled and commissioned at this new offline beamline.
\\
Furthermore, it will be used to perform laser spectroscopy of time-focused ion beams, where we aim to investigate possible systematics and to find the optimal compromise between mass-resolving power and optical resolution.

\subsection{Light collection region}
\begin{figure}[!t]
\includegraphics[width=0.8\linewidth]{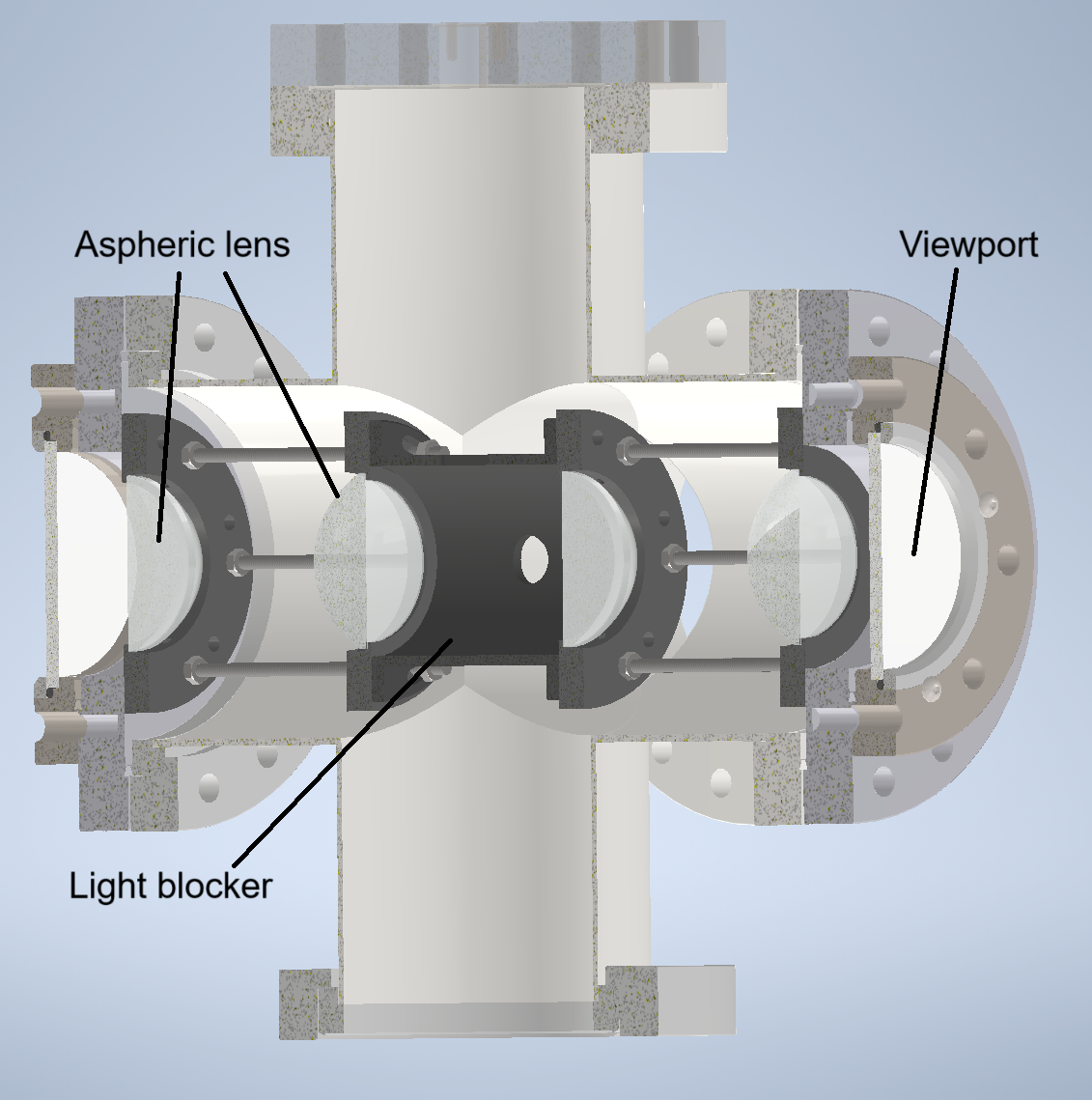}
\caption{\label{fig:LCR} CAD model of the installed LCR chamber. Two aspherical lenses are installed in each sidearm to guide the ions to the PMTs which are placed behind the viewports. In the middle a light blocker is installed to suppress stray light.}
\end{figure}
The light collection region (LCR), illustrated in Fig.\,\ref{fig:LCR}, is based on the design developed at IGISOL \cite{koszorus2023}. The LCR consists of a six-way cross, with two opposing ports equipped with a lens doublet consisting of two aspherical lenses. The diameters of the lenses (B270 Optical Crown Glas) are 50 mm and they have a focal length of 40 mm (ACL5040U from Thorlabs) and 60 mm (AFL50-60-P-U-285 from Asphericon), respectively. The lenses are used to collect and direct the photons towards the photomultiplier tubes (PMT) positioned at either end of the chamber. The optimal positions of the lenses and the PMTs were determined using the Raosi ray-tracing Python package \cite{Gins2022}. To reduce background signal arising from scattered laser light, the interior surfaces of the vacuum chamber and lens holders were coated with Aquadag graphite from Agar scientific. The Aqudag coating was applied via airbrushing with simultaneous heating to ensure uniform coverage and optimal surface resistivity. Additionally, a cylindrical light-blocking tube was installed perpendicular to the laser and ion beam axis. This tube includes two aligned apertures that allow the ion and laser beam to pass through. Its inner surface is coated with Spectral Black foil (Acktar Advanced Coatings) to further suppress stray light.

\subsection{Linear Paul trap}
\begin{figure}[!t]
\includegraphics[width=1.0\linewidth]{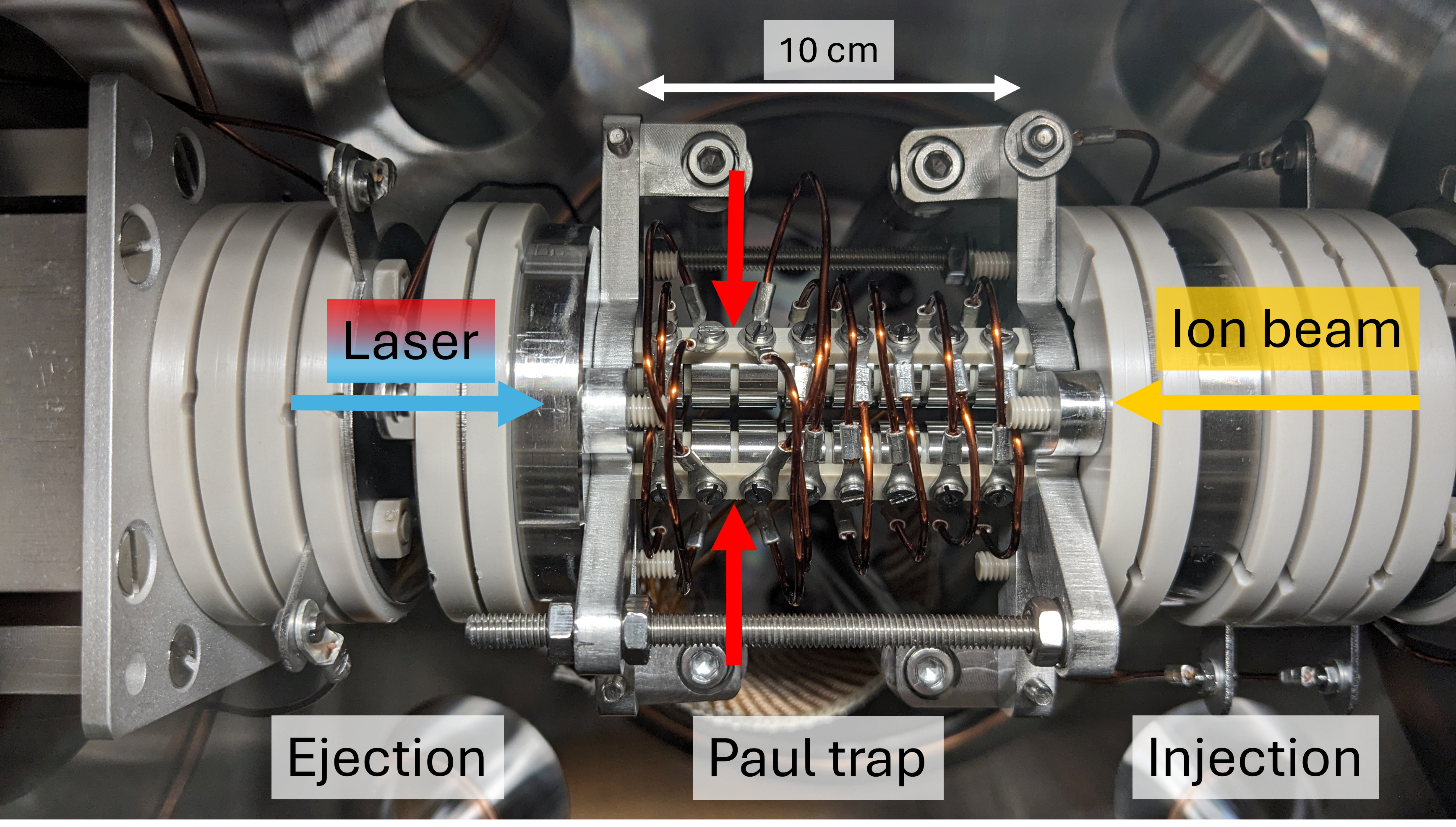}
\caption{\label{fig:trap_photo}Top view of the linear Paul trap, the injection and ejection electrodes inside of the vacuum chamber. After the initial electrostatic deceleration from 10\,keV to roughly 8\,eV, the ions enter the segmented Paul trap with a symmetric radial RF field and a variable axial gradient over eight trap segments. Through fast switching of the potential of the 5th or 6th segment, an axial trap potential with the 8th segment is formed and the ions are trapped in between. Laser access is available in the axial and radial direction to enable laser cooling and spectroscopy of the trapped ions.}
\end{figure}
\begin{figure*}[!t]
\includegraphics[width=1\linewidth]{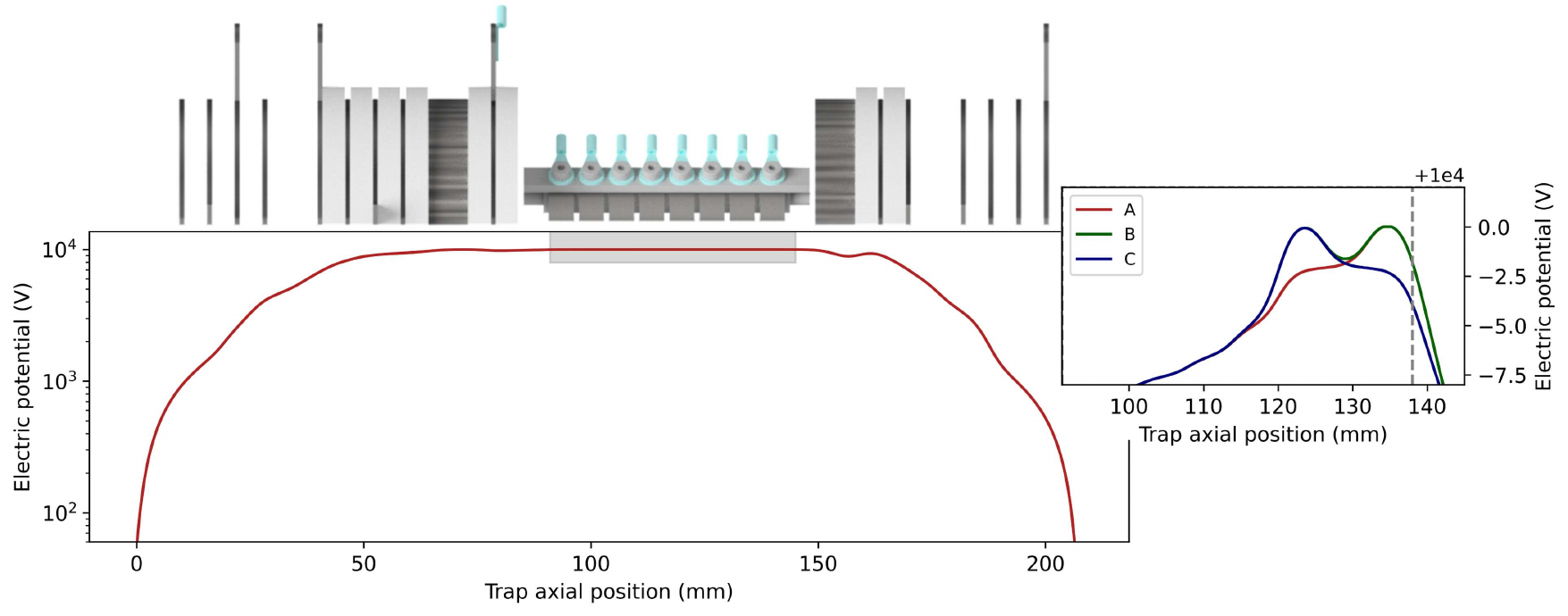}
\caption{\label{fig:trap_voltages}Electric potentials used to decelerate and steer an ion beam with 10\,keV kinetic energy towards the trap, simulated in SIMION, using a realistic model of the ion trap and applying the actual voltages that had resulted in the highest trapping efficiency. The inset shows the difference in voltages during three different trapping stages: A) Prior to ion injection, B) during ion trapping and C) after ion release. The dashed line marks the end of the ion trap.}
\end{figure*}
The linear Paul trap is the central part of STRIPE which aims at hyperfine structure measurements using laser-radiofrequency double-resonance methods, applied to trapped radioactive isotopes produced at RIB facilities \cite{deGroote2024}. Since these ions are typically extracted with a kinetic energy of 10-60\,keV, they have to be decelerated to a few eV to trap them in a segmented Paul trap. We decided on an electrostatic deceleration similar to cooler and buncher devices \cite{Nieminen2001, Mané2009, Schwarz2016, Valverde2020}. In order to achieve necessary trapping times of tens of seconds enabling precision experiments, the trapped ions must be cooled from their initial temperature of typically a few thousand K to <\,mK. This will be achieved by applying laser cooling to the trapped ions. Although this has already been demonstrated within a room-temperature gas-filled cooler and buncher \cite{Sels2022}, our method does not require any buffer gas, improving optical access to the trap and reducing the overall complexity of the setup.
\\
For a first proof-of-principle experiment, the setup was designed for an ion beam energy of 10\,keV/$q$. A detailed manuscript of the full setup including simulations, descriptions of the mechanical and electrical design, and detailed commission measurements will be published later. Here, we focus on the general idea and the main properties.
Figure\,\ref{fig:trap_photo} shows the Paul trap and the injection and ejection electrodes. The trap consists of 8 identical symmetrical segments with individual $U_\mathrm{DC}$ voltages. The length of each segment is 5\,mm with 1\,mm distance between the segments. The radius of the rods $r = 4.6$\,mm while the distance radius $r_0 = 3$\,mm. The Paul trap and the necessary electronics are placed on a high-voltage potential that matches the extraction voltage: $U_\mathrm{trap} = U_\mathrm{source} - 8\,\mathrm{V}$. This decelerates the incoming ions to approximately 8\,eV plus their temperature-dependent initial kinetic energy from the ion source. To achieve a smoother deceleration gradient, several deceleration electrodes are used. These are placed with a 5\,mm separation using PEEK insulating rings. The vacuum chamber which holds the ion trap and decelerators is itself kept at ground potential. The device has been tested up to HVs of 10\,kV without any discharge between the HV and ground. Figure\,\ref{fig:trap_voltages} shows the potential energy, based on SIMION \cite{APPELHANS20051} simulations, which the ions experience as they travel through the setup. 
\\
After the ions have traveled through the decelerator and entered the trap, a symmetric radiofrequency (RF) field with an amplitude of $\approx 120$\,V and a frequency of 1.2\,MHz confines the ions radially. Each of the eight trap segments can be floated individually to a potential from -24\,V to +24\,V relative to $U_\mathrm{trap}$. By adjusting the positive gradient on the first four segments of the Paul trap, the ions are further decelerated. Upon reaching the final segments of the trap, the ions can be trapped in the axial direction through a blocking potential at the last segment and a fast switch of the fifth or sixth segment to the same blocking potential. This is shown in the inset of Fig.\,\ref{fig:trap_voltages}. The latter were simulated in SIMION \cite{APPELHANS20051}, using a realistic model of the ion trap and applying the actual voltages that had resulted in the highest trapping efficiency.
\\
The open design of the trap allows for optical access in all three trap axes for laser cooling, spectroscopy, and light collection. After trapping and potential laser interaction, the ions can also be ejected from the trap, where they can be counted using the particle detectors after the trap.
\subsection{Laser systems}
For the commissioning of the collinear beamline and the Paul trap we have chosen Sr$^+$ ions since they are easy to produce and have a simple and well-known atomic structure. The $5s\,^2\mathrm{S}_{1/2} \rightarrow 5p\,^{2}\mathrm{P}_{1/2}$ (D1-like) transition is usually used as the cooling transition and requires a laser wavelength of 421.5\,nm. Additionally, a 1091.5\,nm-laser is required to repump the ions due to a branching of 5.5\,\% from the $5p\,^2\mathrm{P}_{1/2}$ state to the $4d\,^2\mathrm{D}_{3/2}$ state. We cover the two necessary wavelengths with two Toptica diode lasers which are stabilized to a Fizeau wavemeter (MOGLabs). Although this wavemeter offers only a mediocre accuracy of $\approx 200$\,MHz, its precision is sufficient to address the two transitions. Recalibration of the center wavelengths of these transitions is periodically needed, to compensate for wavemeter readout drifts. Both laser beams are fiber-coupled and transported from the separate laser table to setup where they enter the vacuum chambers of STRIPE or REBEL in free space through an optical viewport.
\section{First results}
\label{sec:results}
In this section, the first results of the commissioning phase of the new apparatus are presented. We demonstrate the current mass-resolving power of the MR-ToF device in Sec.\,\ref{subsec:results_mrtof} and the ion trapping including laser cooling in our linear Paul trap in Sec.\,\ref{subsec:results_trapping}. 
Both experiments use the same surface ion source. In order to discuss our results here also in the possible context of RIB facilities, we investigated the characteristics of the ion source in Sec.\,\ref{subsec:results_ion_energy}.

\subsection{Ion source characteristics}
\label{subsec:results_ion_energy}
\begin{figure}[!tb]
    \centering
    \includegraphics[width=1\linewidth]{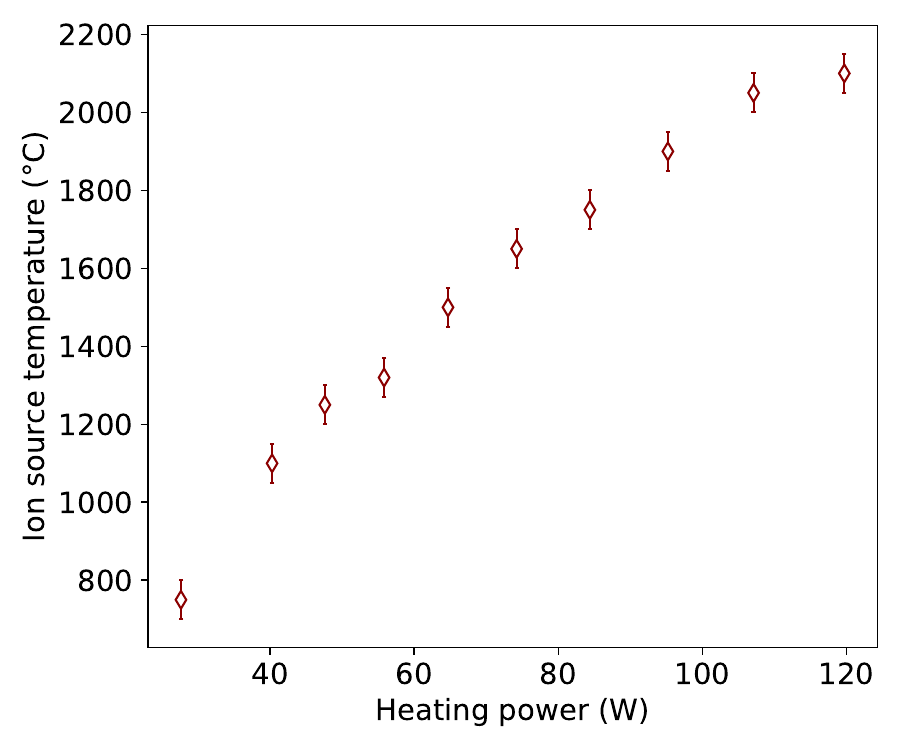}
    \caption{Measured ion source temperature for different heating powers delivered by the current supply. The temperature starts to saturate at 2100\,\degree C for more than 110\,W heating power, which is sufficient to ionize a small fraction ($\approx 0.4$\%) of the Sr vapor in the graphite crucible.}
    \label{fig:ion_source_temp}
\end{figure}

\begin{figure}[!tb]
    \centering
    \includegraphics[width=1\linewidth]{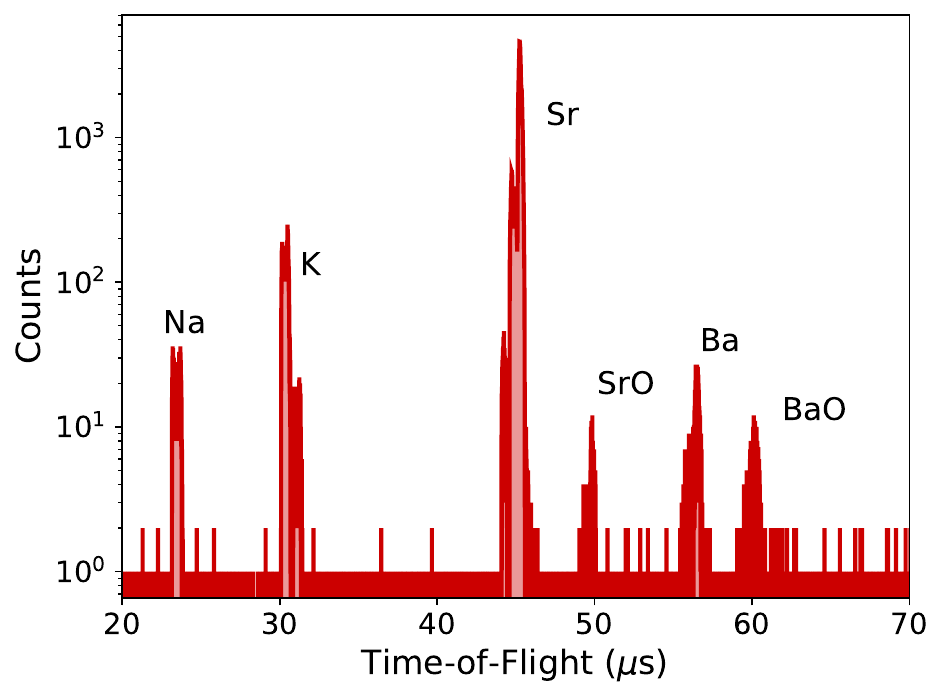}
    \caption{A time-of-flight spectrum recorded at the DS behind the MR-ToF reveals the produced ions in the surface ion source with a heating power of 67\,W which equals to a temperature of approximately 1500\,\degree C. Besides the element of interest (Sr$^+$), small quantities of other elements and molecules are ejected as well. These contaminants are introduced during the handling and filling procedure of the ion source, although care is taken to ensure clean handling.}
    \label{fig:produced_ions}
\end{figure}

The surface ion source is resistively heated by a current supply as explained in Sec.\,\ref{sec:ion_source}. During the first commissioning of the ion source, the temperature of the crucible was measured with a pyrometer for different heating powers. The measurement was stopped at 120\,W heating power to avoid damage to any parts. The recorded data points are plotted in Fig.\,\ref{fig:ion_source_temp}. It is visible that the temperature starts to saturate around 110\,W at a temperature of $T \approx  2100$\,\degree C. From the Saha-Langmuir equation\cite{Copley1935}
\begin{equation}
    \frac{n_+}{n_0} = \frac{g_+}{g_0} \exp \left( \frac{\Phi - E_\mathrm{IP}}{k_\mathrm{B}T} \right),
\end{equation}
the ratio between ions and atoms $n_+/n_0$ can be estimated from statistical weights of ions $g_+$ and atoms $g_0$, the ionization energy $E_\mathrm{IP}$ of the atoms, the work function $\Phi$ of the furnace material, the temperature $T$, and the Boltzmann constant $k_\mathrm{B}$. Since we use a graphite crucible ($\Phi = 4.7$\,eV) to produce Sr ions [$E_\mathrm{IP} (\mathrm{Sr) = 5.7}$\,eV], we can expect an ion-to-atom ratio of $n_+/n_0 \approx 0.4$\% assuming that the strontium has the same temperature as the crucible. However, this is still enough to produce nanoamps of ion beam.
\\
Despite careful handling, unavoidable contamination during the ion source's filling and handling process often results in the ionization of not only Sr but also additional atoms and molecules. Figure\,\ref{fig:produced_ions} shows a time-of-flight spectrum of ejected ion bunches recorded at DS2. In addition to the targeted Sr$^+$ ions, small quantities of other alkaline and earth alkaline elements and their respective oxides are also ejected. However, because of the large mass differences of the contaminants, they quickly separate during the flight path towards STRIPE and REBEL, and they do not disturb the planned measurements.
\begin{figure}[!tb]
\includegraphics[width=1\linewidth]{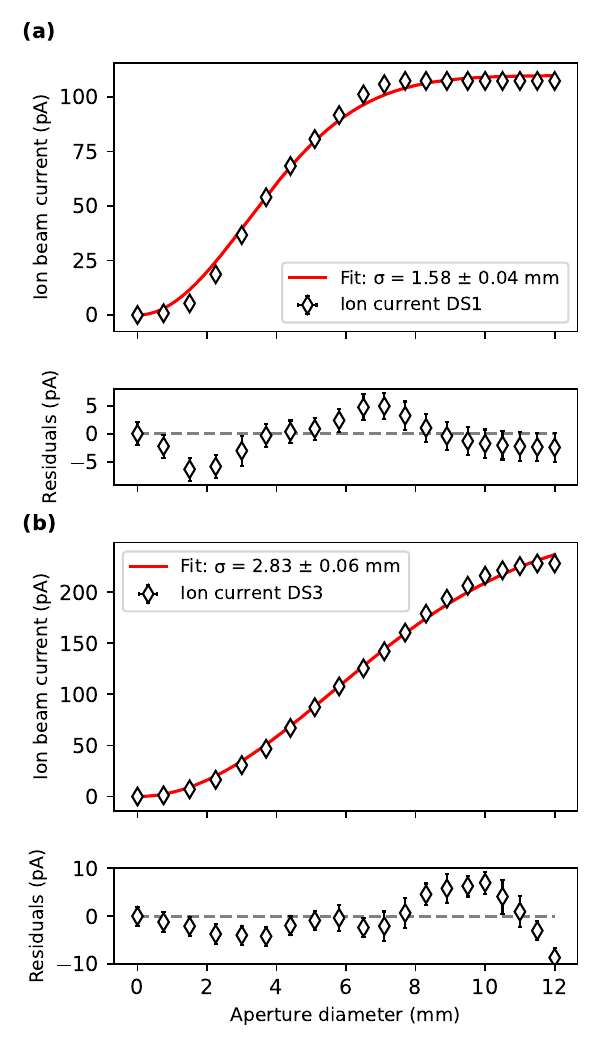}
\caption{Ion beam current vs. the diameter of the iris aperture at DS1 (a) and DS3 (b). The intensity profile of a 2D-Gaussian distribution was fitted to the data points yielding a Gaussian $\sigma$ width of 1.54(3)\,mm and 2.78(6)\,mm. The reduced $\chi^2$ values of 1.5 (a) and 1.7 (b) lead to the conclusion that the beam profiles are not perfectly symmetric, but enough to use the deduced size as a good approximation. \label{fig:IRIS_meas}}
\end{figure}
\\
We estimate the size and shape of the beam by measuring the transmitted ion current for different diameters of the iris diaphragm at DS1 and DS3. Although we cannot derive the two-dimensional beam profile from it, we can assume a symmetric Gaussian beam and fit the expected intensity profile to the measured data, which is given by the integration of a symmetric two-dimensional Gaussian function. The measured data together with the fits is shown in Fig.\,\ref{fig:IRIS_meas}(a) and (b) for DS1 and DS3 respectively. As reflected in the reduced $\chi^2$ values of 1.5 and 1.7 for REBEL and STRIPE, respectively, the assumption agrees reasonably well, but it is not perfect. This means that it is well possible that the beam shape is not exactly symmetric. In addition to this, we assume for the fit function that the center of the ion beam and of the iris aperture perfectly align, but this is not necessarily the case in reality. This can also alter the measurement. However, the fits yield a Gaussian $\sigma$ of the beam size of $\sigma_\mathrm{REBEL} = 1.54(3)$\,mm and $\sigma_\mathrm{STRIPE} = 2.78(6)$\,mm for REBEL and STRIPE, respectively, which we expect to be a good approximation of the beam size.

\subsection{Time-of-flight mass separation}
\label{subsec:results_mrtof}
\begin{figure}[!h]
\includegraphics[width=1\linewidth]{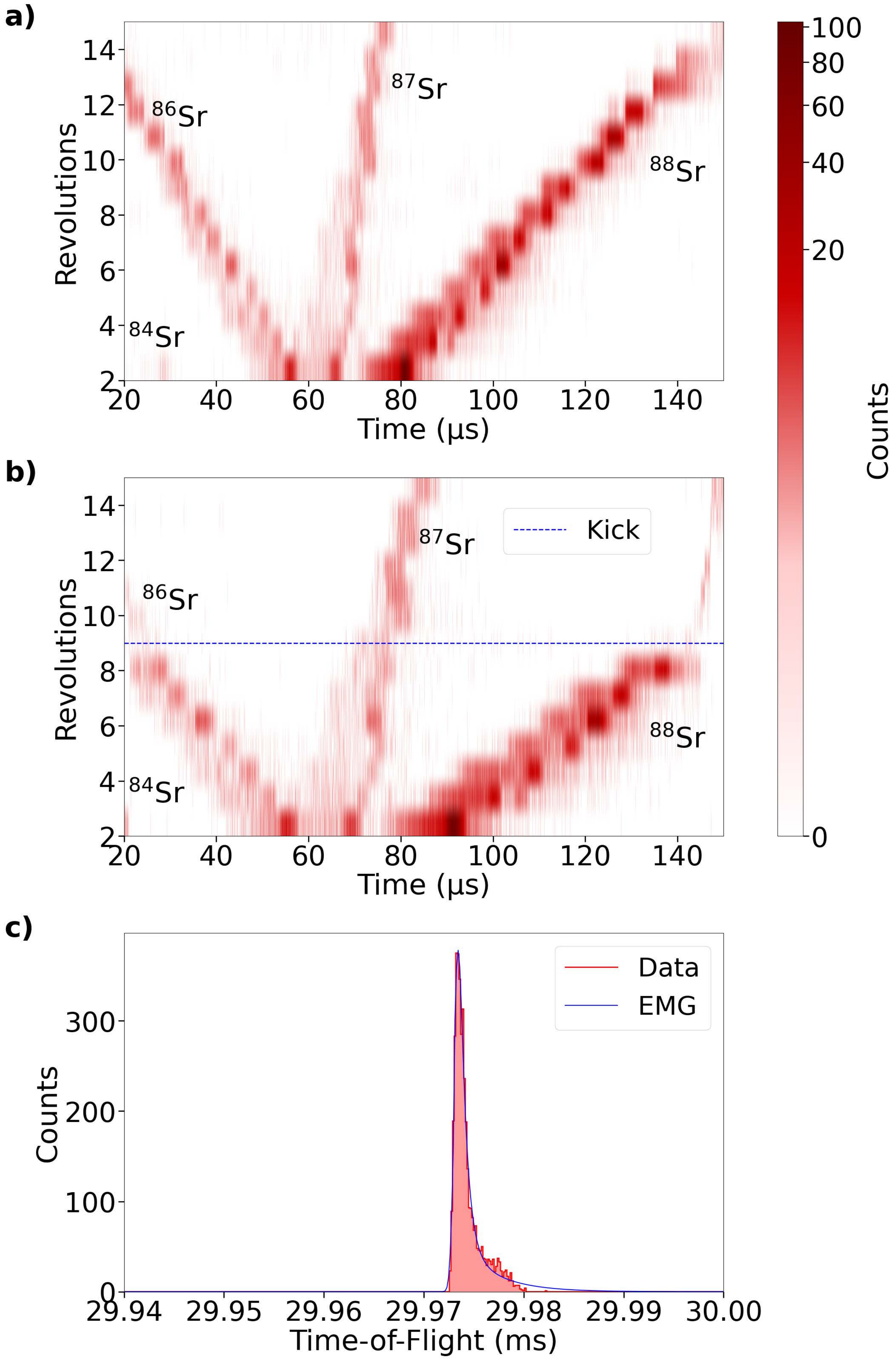}
\caption{\label{fig:results_mrtof}(a) Heatmap plot of several time-of-flight spectra for a different amount of revolutions in the MR-ToF. The separation of the naturally abundant Sr isotopes is clearly visible. (b) Heatmap plot of several revolutions, with the kicker activated at revolution nine (indicated by dashed blue line), a sharp decline in the counts of $^{88}$Sr and $^{86}$Sr is observed. (c) Time-of-flight spectrum for 1000 revolutions of the $^{88}$Sr bunch (red histogram). Together with a fit of a exponentially modified Gaussian (EMG) distribution to the data (blue curve), a mass resolving power of $R = 12855(151)$ can be determined.}
\end{figure}

The MR-ToF device was commissioned with Sr$^+$, which has four naturally abundant isotopes $^{84, 86- 88}$Sr. The goal was to separate these four isotopes in time after the MR-ToF. Although this is a rather easy task because of the large mass difference in comparison to the typical mass resolution of such devices, it is still a good commissioning test for the full apparatus.
\\
The ion beam was tuned from the ion source through the MR-ToF, with the iris apertures in front and behind the MR-ToF set to a diameter of 3\,mm and 12\,mm respectively. With this, we achieved the best mass-resolving power and the best signal-to-background ratio. The overall transmission efficiency from the source to DS2 was approximately 30\,\% with these settings. Time-of-flight spectra were recorded for different numbers of revolutions with a beam energy of 3\,keV. Figure\,\ref{fig:results_mrtof}(a) shows the heatmap of this procedure. The traces of the individual isotopes are clearly visible, and the relative intensities match the natural abundance of Sr. It also shows that those isotopes are indeed easily separable already after a few revolutions.
\\
In order to investigate the mass-resolving power, bunches of $^{88}$Sr$^+$ were trapped for 1000 revolutions, which corresponds to a time-of-flight of approximately 30\,ms. The corresponding spectrum is depicted in Fig.\,\ref{fig:results_mrtof}(c). An exponentially modified Gaussian (EMG)\cite{Lan2001} was fitted to the peak and yields a center value $t_\mathrm{TOF} = 29.973048(3)$\,ms and a FWHM of $\Delta t_\mathrm{bunch} = 1165.9(13)$\,ns. From this, the mass-resolving power
\begin{equation}
    R = \frac{m}{\Delta m} = \frac{t_\mathrm{TOF}}{2 \Delta t_\mathrm{bunch}} \approx 12855 (151)
\end{equation}
can be determined. The main limitation to the resolving power is the energy distribution of the ion bunches due to chopping with the deflection plate.
\\
Finally, the kicking functionality of the MR-ToF was commissioned. Here, unwanted isotopes are removed from the circulating ion bunches by pulsing a kicker electrode with an appropriate timing to deflect the unwanted species. We managed to remove $\geq 95$\% of unwanted strontium isotopes ($^{86}$Sr and $^{88}$Sr) while keeping 92\% of the targeted isotope ($^{87}$Sr) with this method as illustrated in Fig.\,\ref{fig:results_mrtof}(b). This will allow us to investigate the laser spectroscopic properties of such purified ion beams.
\subsection{Ion trapping and laser cooling}
\label{subsec:results_trapping}
\begin{figure}[!tb]
\includegraphics[width=1\linewidth]{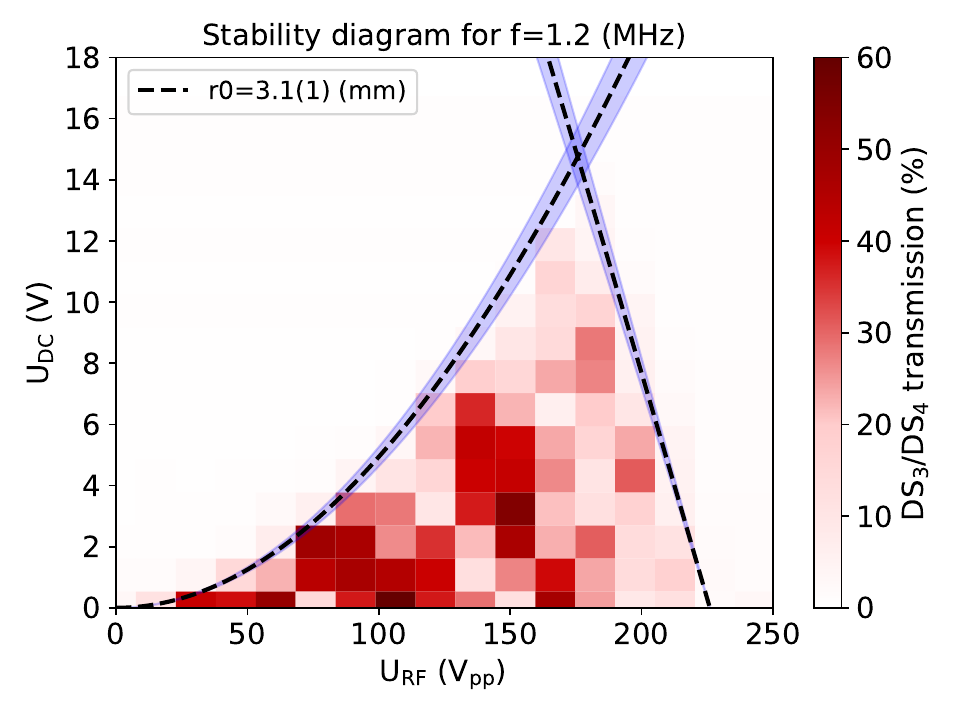}
\caption{\label{fig:results_stability}Stability diagram of the STRIPE trap for an RF frequency of 1.2\,MHz. The measured points shown in a red scale is the transmission of the ions through STRIPE for several peak-to-peak RF voltages (x-axis) and DC voltages (y-axis). The black dashed lines show the theoretical stability lines for our trap radius $r_0 = 3.1(1)$\,mm where the blue shaded area illustrates the given uncertainty.}
\end{figure}

A transmission efficiency of 70\,\% between the ion source and the entrance of the STRIPE setup is typically achieved, using iris apertures set to an opening diameter of 8\,mm. This number is calculated as the ratio of the readouts of the first FC to the FC before the STRIPE setup. The ion trap is operated at a fixed RF frequency $f = 1.2$\,MHz, which is the maximum of the amplifier. The offset voltage between the trap and the ion source was set to -8\,V so that the ions enter the trap with a residual longitudinal kinetic energy of approximately 8\,eV. Multiple combinations of peak-to-peak RF voltages $U_\mathrm{RF}$ and DC offset voltages $U_\mathrm{DC}$ were used, resulting in the heatmap shown in Fig.\,\ref{fig:results_stability}. The agreement with the theoretical Mathieu stability diagram of an oscillating quadrupole field, also shown in the figure using the black lines, is excellent. The best transmission of approximately 55\,\% through STRIPE was achieved with $U_\mathrm{DC} = 3$\,V and $U_\mathrm{RF} = 150$\,V$_\mathrm{pp}$ in transmission mode. This results in a total transport efficiency from the source to DS4 of approximately 38\,\%. In theory, it should be possible to achieve 100\% transmission through a quadrupole ion guide within the stability lines. However, as observed in the past, higher-order components in the trapping potential lead to unstable sections inside of the theoretical stability region of a pure QP field, especially with an ion cloud \cite{Alheit1995}. We expect this effect to be more pronounced in our trap than in other traps due to our non-ideal ratio of $r/r_0 \approx 1.53$, where $r$ is the radius of the rods and $r_0$ is the radius of the inscribed circle tangential to the inner surface of the electrodes. The best QP field approximation in a linear Paul trap is achieved with a ratio of $r/r_0 \approx 1.12$ as derived by Gibson \textit{et al.} \cite{Gibson2001}.
\\
\begin{figure*}[!t]
\includegraphics[width=1\linewidth]{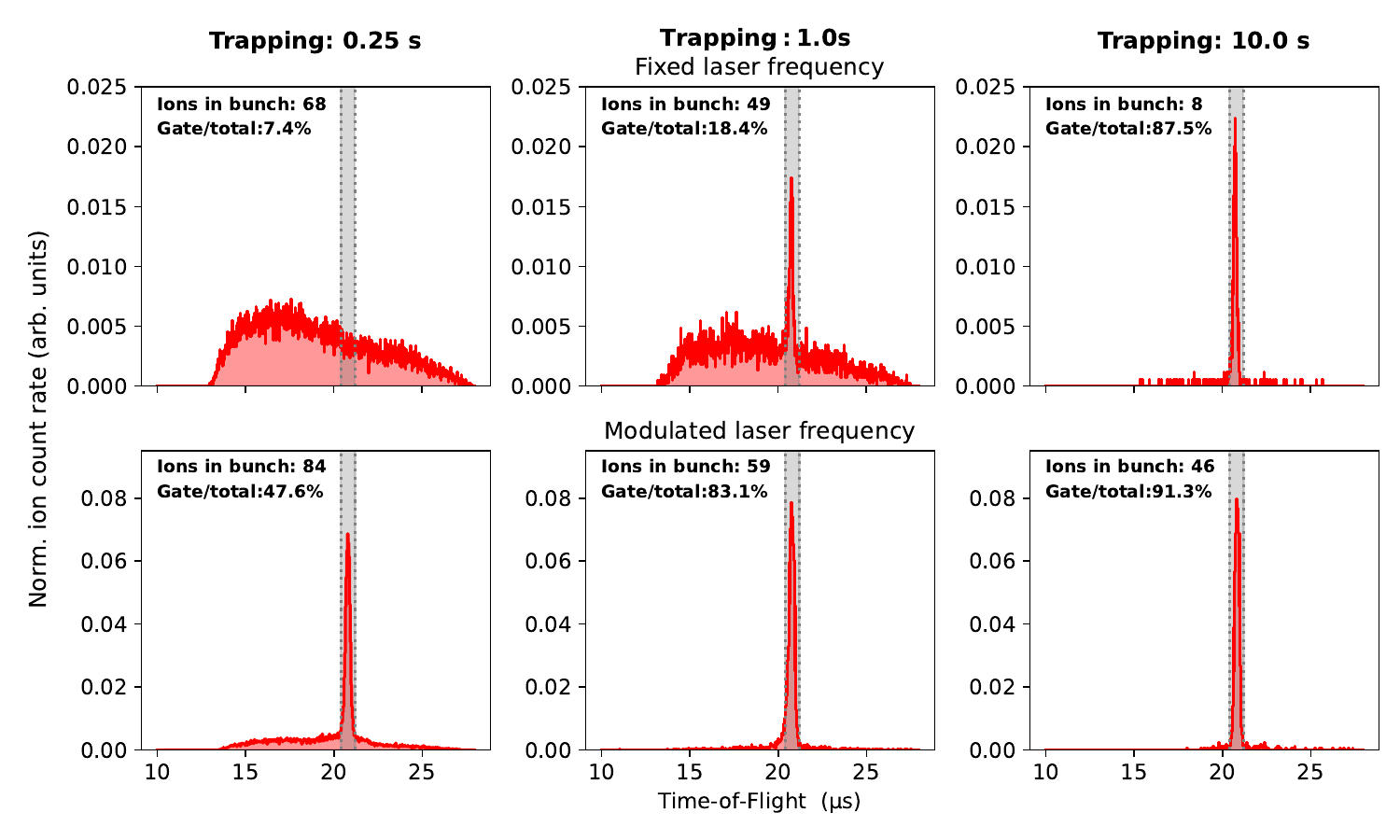}
\caption{\label{fig:laser_cooling_comp}Time-of-flight spectra of ejected ions after laser cooling with a fixed laser frequency (first row) and modulated laser frequency (second row) for 0.25\,s, 1\,s and 10\,s. A narrow peak emerges in both cases which is a clear indication of laser cooling. The modulation of the laser frequency vastly improves the cooling efficiency and speed.}
\end{figure*}
After optimization of transmission, incoming ion bunches of 6\,$\mu$s temporal length were trapped by forming an axial potential well of $U_\mathrm{axial} = 1.8$\,V between the sixth and eighth trap segment. After the ions have been trapped for the targeted amount of time, they were released by lowering the potential of the last electrode and subsequently detected on a MagneTOF detector. By simply confining the decelerated ions without any cooling applied, a trapping efficiency of approximately 40\% of the otherwise transmitted ions was achieved, which adds to a total deceleration and trapping efficiency starting from the ion source of approximately 15\%. Part of the losses may be due to the spatial distribution and energy spread caused by the electrostatic chopping; better results could thus be expected when re-trapping ion bunches extracted from typical gas-filled cooler-bunchers which are present in most RIB laboratories. The amount of trapped ions has been varied from single ions to approximately 100 ions per bunch, yielding similar trapping and transmission efficiencies, as well as similar storage half-lives of the trapped ions (on the order of 2.2(2)\,s). We assume that the main loss channel here is collisional heating and/or molecular formation at the comparably high residual gas pressure of $p_\mathrm{trap} = 9\times10^{-9}$\,mbar.
\\
\subsubsection{Laser cooling tests}
By sending a 422\,nm-laser along the axial trap direction, and a 1092\,nm-laser in the radial trap direction, the $5s\,^2\mathrm{S}_{1/2} \rightarrow 5p\,^{2}\mathrm{P}_{1/2}$ transition and the $4d\,^2\mathrm{D}_{3/2} \rightarrow 5p\,^{2}\mathrm{P}_{1/2}$ transition in $^{88}$Sr can be addressed. The infrared laser was tuned to the resonance frequency, while the ultraviolet laser was set to a fixed frequency 20\,MHz red-detuned from the resonance frequency to apply an effective cooling force. For optimal cooling speed, the 1092\,nm-laser was generally configured at 1\,mW, whereas the 422\,nm-laser was adjusted to 20\,mW. The first row of Fig.\,\ref{fig:laser_cooling_comp} shows the ToF spectra of the ejected ions for a trapping time of 0.25\,s, 1\,s and 10\,s. While after 250\,ms no effect is visible in comparison to the uncooled bunch, a clear peak structure with a reduced TOF width becomes apparent after 1\,s of cooling. Because of the reduced kinetic energy of the cooled ions, their spatial extension in the axial potential is reduced, which leads to a compressed time structure in the ejected bunch. The fraction of the ions which is laser-cooled continues to increase with time, balanced by the trapping half-life of the ions. Indeed, Fig.\,\ref{fig:laser_cooling_comp} also shows that the total number of trapped ions is strongly reduced, as many are lost prior to being cooled. The ions which are cooled, however, have a substantially longer trapping half-life,  approximately 20\,s. 
\\
The reason for the slow timescale of the cooling process is the small laser linewidth of the blue laser compared to the large Doppler broadening of the trapped ions. Thereby, the ions can interact only for a very short time with the laser when their velocity is close to zero at the turning points of their trajectory, which increases the total amount of time for the laser-cooling process. This can be resolved by ramping the frequency of the cooling laser, starting from a far red-detuned frequency. This leads to an effective compression of the ion temperature because the hottest ions are cooled first and then cooled together with slower ions. This ramp was obtained by tuning the piezo voltage of the external grating of the laser using a function generator (Keysight 33500B). The best parameters were obtained using a laser frequency range of 1.5\,GHz, swept over in a 370\,ms time span. The second row in Fig.\,\ref{fig:laser_cooling_comp} illustrates the improvement in cooling time achieved with this frequency ramp. Already after 0.25\,s roughly 48\% of the trapped ions are laser cooled, and after 1\,s the ratio is increased to 83\%. Additionally, significantly less ions are lost from the start, resulting in a more than five times higher number of trapped ions after 10\,s than in the fixed-frequency mode. We tried different ramping functions, such as a simple linear or exponential one, but the differences in cooling efficiency and speed were minor. The full efficiency from the ion source to laser-cooled ions after 10\,s trapping time is then roughly 4\%.
\section{Conclusion and future plans}\label{sec:conclusion}
We presented a new offline beamline for the development of new techniques in laser spectroscopy of radioactive isotopes. While the REBEL part is devoted to the extension and improvement of established collinear laser spectroscopy and MR-ToF mass separators, the STRIPE part aims for the transfer of existing high-precision ion trap technology into the context of radioactive beam facilities. First commissioning and proof-of-principle measurements have been performed in both parts of the beamline.
\\
For REBEL, the successful mass separation of naturally abundant Sr isotopes with more than 600 revolutions in an MR-ToF represents an important milestone towards laser spectroscopy of a mass-separated ion beam. The combination of an MR-ToF and CLS or CRIS will enable the investigation of more exotic nuclei that have not been measured before due to isobaric contamination. The next steps include the installation and commissioning of the already designed LCR and finally CLS of mass-separated ion bunches from a MR-ToF device. This will give important information on the MR-ToF properties and the longitudinal energy spread of the ion bunches through the observed spectroscopic lineshape. Furthermore, the replacement of the kicker electrode with a Bradburry-Nielsen gate similar to the design of Wolf \textit{et al.} \cite{Wolf2012} will improve the resolving power and is expected to also improve the transmission efficiency.
\\
The successful demonstration of the deceleration, stopping, and laser cooling of ions with a kinetic energy of 10\,keV with STRIPE is an important achievement. Upcoming upgrades will include the installation of a light collection system for the trapped ions. This will give us more information about the temperature of the ions and enable the possibility for laser spectroscopy, also using dipole-forbidden lines. In order to address those transitions, we plan to acquire an ultra-stable high-finesse cavity for laser stabilization to achieve laser linewidths on sub-kHz level.
\\
We expect that the overall trapping efficiency can still be improved by loading the trap from a conventional He-filled cooler and buncher instead of a surface ion source directly, and by optimizing the trap dimensions as well as the laser-cooling process by modulating the laser frequency in a more controlled way with a broadband traveling-wave electro-optic modulator. With this, a fast scanning sideband could be modulated onto the far red-detuned carrier frequency.
\\
However, even with the current performance, there are promising opportunities to use this trap at RIB facilities. As simulated in Ref.\,\onlinecite{Sels2022}, STRIPE can already be used as a cooler and buncher device to greatly improve the precision of PI-ICR mass measurements \cite{Eliseev2014}. Preparations for a first proof-of-principle experiment with JYFLTRAP at IGISOL \cite{Nesterenko2018} are already ongoing. Furthermore, if care is taken to compensate for possible external magnetic fields, this trap could be considered as a means to deliver to spin-polarized/aligned ensembles to other experiments.
\\
Potential candidates for the first laser spectroscopy of trapped radioactive ions are listed in Tab.\,\ref{tab:candidates}. Those are species which are well-established in trapped-ion studies since they are directly laser-coolable with only one or two lasers, which would make them ideal first targets. The isotopes listed in Tab.\,\ref{tab:candidates} have a lifetime with more than a second, which is sufficient time for cooling and measurement, and a yield of more than 100 ions per second, ensuring that at least one ion can be loaded per second given our current overall efficiency. Tab.\,\ref{tab:candidates} lists values from the ISOLDE yield database as a reference. Addressing clock transitions in these species over long isotope chains can provide high accuracy data valuable for nuclear structure studies, and may also add to the ongoing search for physics beyond the standard model \cite{Berengut2020}.
\begin{table}[!tb]
    \caption{Potential candidates for the first laser spectroscopy of trapped radioactive ions. All listed isotopes have a lifetime of more than a second and an ISOLDE yield of more than 100 ions per second. This reflects our cooling time and our overall transport efficiency.}
    \centering
    \begin{tabular}{c|c|c|c}
        ~ & Isotopes $A$ & \shortstack{Lifetime \\ / s} & \shortstack{ISOLDE \\ Yield / s} \\\hline
         Be$^+$& 7, 10, 11 & $\geq 13.77$& $\geq 7\times 10^6$ \\
         Mg$^+$& 22, 23, 27-29 & $\geq 1.3$& $\geq 1.6\times 10^6$ \\
         Ca$^+$& 41, 45, 47, 49-52 & $\geq 4.6$& $\geq 1\times 10^2$ \\
         Sr$^+$& 76 - 83, 85, 89 - 96& $\geq 1$& $\geq 8\times 10^2$ \\
         Cd$^+$& 99 - 105, 107, 109, 115, 117 - 124 & $\geq 1.25$& $\geq 5\times 10^2$ \\
         Ba$^+$ & 116 - 129, 131, 133, 139 - 146 & $\geq 1.3$& $\geq 1.9\times 10^5$ \\
         Yb$^+$& 153, 155 - 167, 169, 175, 177 - 180 & $\geq 1.7$& $\geq 2.5\times 10^4$ \\
         Ra$^+$& 207 - 214, 221 - 232 & $\geq 1.3$& $\geq 5.6\times 10^3$ \\
    \end{tabular}
    \label{tab:candidates}
\end{table}
\\
To overcome the restriction of directly laser-coolable species, we intend to investigate the sympathetic cooling with STRIPE, whereby ions of a laser-coolable species will be co-injected with the ions of interest. Here, we will investigate the time scales of this cooling mechanism, the efficiency, and possible ion load limitations, in order to establish its feasibility for radioactive isotopes. This would allow us to produce cold samples of almost every radioactive element produced in a RIB facility with STRIPE.

\begin{acknowledgments}
We are grateful for the strong support during the initial commissioning of the new beamline, especially the help of Frank Wienholtz and the MR-ToF Collaboration was very valuable and saved a lot of time during the installation of the MR-ToF. Furthermore, we wish to acknowledge the support from Kristian K\"onig and Wilfried N\"ortersh\"auser by providing the high-voltage divider and the design of the ion source and by sharing some HV modules. We also thank Wouter Gins and the IGISOL collaboration, as well as Kieran Flanagan and the CRIS collaboration, for sharing the design of the QP triplet and the switchyard. The contribution of the mechanical and electrical workshop of the Instituut voor Kern- en Stalingsfysica (KU Leuven) is appreciated as well.
\\
The installation of this setup was supported by KU Leuven (start-up grant STG/23/030; C14/22/104 and annex C+/24/005) and Fonds Wetenschappelijk Onderzoek (projects G0A0723N, G053925N, G080022N, G081422N; Odysseus project G0F7321N; IRI funding I001323N).
\end{acknowledgments}

\section*{Author declarations}
\subsection*{Conflict of Interest}
The authors have no conflicts to disclose.

\subsection*{Author Contributions}
\textbf{Phillip Imgram}: Conceptualization (supporting); Formal analysis (supporting); Investigation (equal); Methodology (lead); Project administration (lead); Software (supporting); Visualization (equal); Writing – original draft (lead); Writing – review \& editing (equal).
\textbf{Dinko Atanasov}: Investigation (supporting), Validation (supporting); Writing – review \& editing (equal).
\textbf{Michail Athanasakis-Kaklamanakis}: Conceptualization (supporting), Writing – review \& editing (equal).
\textbf{Paul Van den Bergh}: Investigation (supporting), Methodology (supporting), Software (supporting), Validation (supporting); Writing – review \& editing (equal).
\textbf{Tobias Christen}: Formal analysis (equal); Investigation (equal); Methodology (supporting); Visualization (equal); Writing – review \& editing (equal).
\textbf{Ruben de Groote}: Conceptualization (equal); Funding acquisition (equal), Methodology (supporting); Project administration (supporting); Software (supporting); Supervision (equal); Validation (equal); Writing – original draft (support); Writing – review \& editing (equal). 
\textbf{\'Agota Koszor\'us}: Conceptualization (equal); Funding acquisition (equal); Methodology (supporting); Project administration (supporting); Supervision (equal); Validation (equal); Writing – review \& editing (equal).
\textbf{Gerda Neyens}: Conceptualization (supporting); Funding acquisition (supporting); Supervision (supporting); Writing – review \& editing (equal).
\textbf{Stefanos Pelonis}: Formal analysis (equal); Investigation (equal), Methodology (equal), Software (lead); Visualization (equal); Writing – review \& editing (equal).

\section*{Data Availability Statement}
The data that support the findings of this study are available from the corresponding author upon reasonable request.

\bibliography{mybib}

\end{document}